\documentclass[12pt]{article}
\usepackage{amssymb}

\newcommand{\alphab}{\mbox{\boldmath $\alpha$}}
\newcommand{\ab}{\mbox{\boldmath $a$}}
\newcommand{\xib}{\mbox{\boldmath $\xi$}}
\newcommand{\bb}{\mbox{\boldmath $b$}}
\newcommand{\tv}{\tilde{V}}
\newcommand{\lambdab}{\mbox{\boldmath $\lambda$}}
\newcommand{\tx}{\tilde{x}}
\def\case#1#2{{\textstyle{\frac{#1}{#2}}}}
\def\sech{\mathop{\rm sech}\nolimits}
\def\csch{\mathop{\rm csch}\nolimits}

\title{
Hamiltonians with position-dependent mass, deformations and
supersymmetry}

\author{\underline{C.\ Quesne}$^1$,B.\ Bagchi$^2$, A.\ Banerjee$^2$,  V.\ M.\
Tkachuk$^3$\\
{\small $^1$ Physique Nucl\'eaire Th\'eorique et Physique Math\'ematique, 
Universit\'e Libre de Bruxelles,} \\  {\small Campus de la Plaine CP229,
Boulevard~du Triomphe, B-1050 Brussels, Belgium}\\  
{\small $^2$ Department of Applied Mathematics, University of Calcutta,} \\
{\small 92 Acharya Prafulla Chandra Road, Kolkata 700 009, India}\\ 
{\small $^3$ Ivan Franko Lviv National University, Chair of Theoretical
Physics,}\\
{\small 12, Drahomanov Street, Lviv UA-79005, Ukraine}}

\date{}

\begin{document}

\maketitle
\begin{abstract}
A new method for generating exactly solvable Schr\"odinger equations with a
position-dependent mass is proposed. It is based on a relation with
some deformed Schr\"odinger equations, which can be dealt with by using a
supersymmetric quantum mechanical approach combined with a deformed
shape-invariance condition. The solvability of the latter is shown to impose
the form of both the deformed superpotential and the position-dependent
mass. The conditions for the existence of bound states are
determined. A lot of examples are provided and the corresponding bound-state
spectrum and wavefunctions are reviewed
\end{abstract}

%%%%%%%%%%%%%%%%%%%%%%%%%%%%%%%%%%%%%%%%%%%%%%%%%%%%%%%%%%%%%
% The main text of your paper                               %
%%%%%%%%%%%%%%%%%%%%%%%%%%%%%%%%%%%%%%%%%%%%%%%%%%%%%%%%%%%%%

\section{Introduction}

Schr\"odinger equations with a position-dependent mass~(PDM) play an important
role in many physical problems. They appear in the energy-dependent functional
approach to quantum many-body systems~\cite{ring} (e.g., nuclei, quantum
liquids, $^3$He clusters, metal clusters) and are very useful in the
description of electronic properties of condensed-matter
systems~\cite{bastard} (e.g., compositionally-graded crystals, quantum dots,
liquid crystals). The PDM presence may also reflect other unconventional
effects, such as  deformation of the canonical commutation relations or
curvature of the underlying space~\cite{cq04} or else pseudo-Hermiticity of the
Hamiltonian~\cite{jones}.\par
%
%----------------------------------------------------
%
Several exactly solvable, quasi-exactly solvable or conditionally exactly
solvable PDM Schr\"odinger equations have been constructed using point
canonical transformations, Lie algebraic methods or supersymmetric quantum
mechanical (SUSYQM) and shape-invariance (SI) techniques (see \cite{bagchi04,
bagchi05a} and references quoted therein). Most of them can be obtained from
known constant-mass models by changes of variable and of function. As a
consequence the spectrum is left unchanged although the potential is given by
a complicated mass-deformed expression.\par
%
%-----------------------------------------------------
%
Here we present a new method~\cite{bagchi05b}, generalizing the work
in~\cite{cq04}. It is based on a relation with deformed Schr\"odinger equations
and uses a SUSYQM method combined with a deformed SI condition. The resulting
spectra will contrast with those obtained in the constant-mass case.\par
%
%===================================================
%
\section{PDM and deformed Schr\"odinger equations}

As is well known, the noncommutativity of a PDM with the momentum operator
results in an ambiguity in the kinetic energy operator definition coming
from the selected ordering. On choosing the von Roos general two-parameter form
of $T$~\cite{vonroos}, which has an inbuilt Hermiticity and contains all the
plausible forms as special cases, and setting $\hbar = 2m_0 = 1$, the PDM
Schr\"odinger equation can be written as
\begin{equation}
  \left[- \frac{1}{2} \left(M^{\xi'} \frac{d}{dx} M^{\eta'} \frac{d}{dx}
  M^{\zeta'} + M^{\zeta'} \frac{d}{dx} M^{\eta'} \frac{d}{dx} M^{\xi'} \right) 
  + V\right] \psi =E \psi.  \label{eq:PDM-SE}
\end{equation}
Here $V(\ab;x)$ is the potential, $M(\alphab; x)$ is the dimensionless form of
the mass function 
$m(\alphab; x) = m_0 M(\alphab; x)$, $\ab$ and $\alphab$  denote two sets of
parameters, and the von Roos ambiguity parameters $\xi'$, $\eta'$, $\zeta'$
are constrained by the condition
$\xi' + \eta' + \zeta' = -1$.\par
%
%-------------------------------------------------------
%
On setting $M(\alphab; x) = [f(\alphab; x)]^{-2}$ and $f(\alphab; x) =
1 + g(\alphab; x)$, where $f(\alphab; x)$ is some positive-definite function
and $g(\alphab; x) = 0$ corresponds to the constant-mass case,
Eq.~(\ref{eq:PDM-SE}) becomes 
\begin{equation}
  \left[- \frac{1}{2} \left(f^{\xi} \frac{d}{dx} f^{\eta} \frac{d}{dx}
  f^{\zeta} + f^{\zeta} \frac{d}{dx} f^{\eta} \frac{d}{dx} f^{\xi}\right)
  + V\right] \psi = E \psi  \label{eq:PDM-SE-bis}
\end{equation}
with $\xi + \eta + \zeta = 2$. The ambiguity parameters $\xi$, $\eta$, $\zeta$
(denoted collectively by $\xib$) can be transferred from the kinetic energy
term to an effective potential
\begin{equation}
  V_{\rm eff}(\bb; x) = V + \tv, \qquad \tv = \rho f f'' + \sigma f^{\prime2},
  \label{eq:Veff} 
\end{equation}
so that Eq.~(\ref{eq:PDM-SE-bis}) acquires the form
\begin{equation}
  H \psi \equiv \left[- \left(\sqrt{f}\,\frac{d}{dx} \sqrt{f} \right)^2 +
  V_{\rm eff}\right] \psi = E \psi.  \label{eq:PDM-SE-ter}
\end{equation}
In (\ref{eq:Veff}), a prime denotes derivative with respect to $x$,  $\rho =
\frac{1}{2} (1 - \xi - \zeta)$, $\sigma = \left(\frac{1}{2} - \xi\right)
\left(\frac{1}{2} - \zeta\right)$ and $\bb = (\ab, \alphab, \xib)$.
\par
%
%--------------------------------------------------------
%
Equation (\ref{eq:PDM-SE-ter}) can be reinterpreted as a deformed
Schr\"odinger equation with $H = \pi^2 + V_{\rm eff}$, where $\pi = \sqrt{f}\,
p \sqrt{f}$ satisfies the deformed commutation relation $[x,
\pi] = {\rm i} f$ and $f(\alphab;x)$ acts as a deforming function.\par
%
%=======================================================
%
\section{General procedure}

The first step in our procedure consists in taking for $V_{\rm eff}$ any known
SI potential under parameter translation. This means that the initial potential
$V$ will be determined by inverting (\ref{eq:Veff}) as $V = V_{\rm eff} -
\tv$, so that its parameters $\ab$ will be $\ab = (\bb, \alphab, \xib)$.\par
%
%---------------------------------------------------------------------
%  
We then consider $H$ as the first member $H_0 = H$ of a hierarchy of 
Hamiltonians
\begin{equation}
  H_i = A^+(\alphab, \lambdab_i) A^-(\alphab, \lambdab_i) + \sum_{j=0}^i 
  \epsilon_j, \qquad i = 0, 1, 2, \ldots,  \label{eq:H_i} 
\end{equation}
where the first-order operators
\begin{equation}
  A^{\pm}(\alphab, \lambdab_i) = \mp \sqrt{f(\alphab; x)}\,\frac{d}{dx} 
  \sqrt{f(\alphab; x)} + W(\lambdab_i; x)  \label{eq:A}
\end{equation}
satisfy a deformed SI condition
\begin{equation}
  A^-(\alphab, \lambdab_i) A^+(\alphab, \lambdab_i) = A^+(\alphab, 
  \lambdab_{i+1})
  A^-(\alphab, \lambdab_{i+1}) + \epsilon_{i+1}
  \label{eq:deformed-SI}  
\end{equation}
for $i = 0, 1, 2, \ldots$. Here $\epsilon_i$ and $\lambda_i$, $i=0$, 1,
2,~\ldots, are some constants. Solving Eq.~(\ref{eq:deformed-SI}) means
that it is possible to find a superpotential $W(\lambdab; x)$, a deforming
function $f(\alphab; x)$ and some constants
$\lambda_i$, $\epsilon_i$, $i=0$, 1, 2,~\ldots, with $\lambdab_0 =
\lambdab$, such that
\begin{equation}
  V_{\rm eff}(\bb; x) = W^2(\lambdab; x) - f(\alphab; x) W'(\lambdab; x) + \epsilon_0
  \label{eq:C1}
\end{equation}
and 
\begin{eqnarray}
  && W^2(\lambdab_i; x) + f(\alphab; x) W'(\lambdab_i; x) \nonumber \\
  && \mbox{} = W^2(\lambdab_{i+1}; x) - f(\alphab; x) W'(\lambdab_{i+1}; x) +
      \epsilon_{i+1}, \quad i = 0, 1, 2, \ldots.  \label{eq:C2}
\end{eqnarray}
As a consequence, the (deformed) SUSY partner $H_1$ of $H$ will be characterized by a
potential $V_{{\rm eff},1}(\bb, \alphab, \lambdab; x) = V_{\rm eff}(\bb; x) + 
2 f(\alphab; x) W'(\lambdab; x)$. Note that in Eqs.~(\ref{eq:C1}) and
(\ref{eq:C2}), the additional terms with respect to the undeformed case are
proportional to $g W'$.
\par
%
%-----------------------------------------------------------------
% 
To find a solution for $W$, $f$, $\epsilon_i$ and $\lambda_i$, our strategy
consists in (i) assuming that the deformation does not affect the form of
$W$ but only brings about a change in its parameters $\lambdab$ (which will now
also depend on $\alphab$), and (ii) choosing $g(\alphab; x)$ in such a way that
the function $g(\alphab;x) W'(\lambdab; x)$ contains the same kind of terms as
those already present in the undeformed case, i.e., $W^2(\lambdab; x)$ and
$W'(\lambdab; x)$.\par
%
%------------------------------------------------------
%
Once we have found a solution to Eqs.~(\ref{eq:C1}) and (\ref{eq:C2}), the
bound-state energy spectrum and corresponding wavefunctions can be found as in
conventional SUSYQM. They may be written as
\begin{equation}
  E_n(\alphab, \lambdab) = \sum_{i=0}^n \epsilon_i
\end{equation}
and
\begin{equation}
  \psi_n(\alphab, \lambdab; x) \propto \frac{1}{\sqrt{f(\alphab; x)}}\,
   \varphi_n(\alphab, \lambdab; x)\exp\left(- \int^x \frac{W(\lambdab_n; \tx)}
  {f(\alphab; \tx)} d\tx\right),  \label{eq:exc-wf} 
\end{equation}
where $\varphi_n(\alphab, \lambdab; x)$ fulfils the equation
\begin{eqnarray}
  &&\varphi_{n+1}(\alphab, \lambdab; x) = - f(\alphab; x)
\varphi'_n(\alphab,
\lambdab_1;
  x) \nonumber \\ 
  && \mbox{} + [W(\lambdab_{n+1}; x) + W(\lambdab; x)] \varphi_n(\alphab,
\lambdab_1; x)
  \label{eq:phi-eq}
\end{eqnarray}
with $\varphi_0(\alphab, \lambdab; x) = 1$. They remain however formal
solutions till one has checked that they satisfy appropriate physical
conditions. In the PDM or deformed case, such conditions are twofold: (i)
square integrability on the interval of definition of $V_{\rm eff}$,  as in
the conventional case, and (ii) Hermiticity of $\pi$ or, equivalently, of
$H$ in the corresponding Hilbert space. The latter imposes that
$|\psi|^2 f = \psi^2/\sqrt{M}$ vanishes at the end points of the
interval. This extra condition may have some relevant effects whenever
the PDM vanishes there.\par
%
%====================================================
\section{\boldmath Classes of superpotentials and corresponding $g(\alphab;x)$}

One can show that all the known potentials which are SI under parameter
translation fall into three classes. For all of them, we have determined the
general form of the deforming function allowing Eqs.~(\ref{eq:C1}) and
(\ref{eq:C2}) to remain solvable in accordance with the general strategy
reviewed in Sec.~3.\par
%
%----------------------------------------------------
%
In terms of some parameter-independent function $\phi(x)$ and of two parameters
$\lambda$, $\mu$ making up the set $\lambdab$, the results are given by
\begin{equation}
  \begin{array}{ll}
     \mbox{Class 1:} & W(\lambdab; x) = \lambda \phi(x) + \mu,  \\[0.3cm]
     & \phi'(x) = A \phi^2(x) + B \phi(x) + C,  \\[0.3cm]
     & g(\alphab; x) = \frac{\displaystyle A'(\alphab) \phi^2(x) + B'(\alphab)
       \phi(x) + C'(\alphab)}{\displaystyle A \phi^2(x) + B \phi(x) + C}, 
       \\[0.3cm]
     \mbox{Class 2:} & W(\lambdab; x) = \lambda \phi(x) +
       \frac{\displaystyle \mu}{\displaystyle \phi(x)}, 
       \\[0.3cm]
     & \phi'(x) = A \phi^2(x) + B, \\[0.3cm]  
     & g(\alphab; x) = \frac{\displaystyle A'(\alphab) \phi^2(x) +
       B'(\alphab)}{\displaystyle A \phi^2(x) + B},  \\[0.3cm]
     \mbox{Class 3:} & W(\lambdab; x) = \frac{\displaystyle \lambda \phi(x) +
       \mu}{\displaystyle \sqrt{A \phi^2(x) + B}},  \\[0.3cm]
     & \phi'(x) = [C \phi(x) + D] \sqrt{A \phi^2(x) + B}, \\[0.3cm] 
     & g(\alphab; x) = \frac{\displaystyle C'(\alphab) \phi(x) + D'(\alphab)}
       {\displaystyle C \phi(x) + D}. 
  \end{array}   \label{eq:3cases}
\end{equation}
Here $A$, $B$, $C$, $D$ and $A'(\alphab)$, $B'(\alphab)$, $C'(\alphab)$,
$D'(\alphab)$ are some numerical and $\alphab$-dependent constants,
respectively.\par
%
%----------------------------------------------------
%
In all three cases, the integral on the right-hand side of
Eq.~(\ref{eq:exc-wf}) can be explicitly carried out by simple integration
techniques. Furthermore, by changes of variable and of function,
$\varphi_n(\alphab, \lambdab; x)$ in the same equation  can be transformed
into an $n$th-degree polynomial $P_n(\alphab, \lambdab; y)$ as follows: 
\begin{equation}
  \begin{array}{lll}
     \mbox{Class 1:} & \varphi_n(\alphab, \lambdab; x) = P_n(\alphab,
       \lambdab;y), & y = \phi(x),  \\[0.2cm]
     \mbox{Class 2:} & \varphi_n(\alphab, \lambdab; x) = y^{-n/2}
       P_n(\alphab, \lambdab;y), & y = \phi^{-2}(x), \\[0.2cm] 
     \mbox{Class 3:} & \varphi_n(\alphab, \lambdab; x) = (A y^2 + B)^{-n/2}
       P_n(\alphab, \lambdab;y), & y = \phi(x).  \\[0.2cm]
  \end{array}   \label{eq:3casesbis}
\end{equation}
Such polynomials are related to deformed classical orthogonal polynomials
and satisfy the equations:
\begin{equation}
  \begin{array}{ll}
     \mbox{Class 1:} & P_{n+1}(\alphab, \lambdab; y) = - \{[A + A'(\alphab)]
       y^2 \\[0.2cm] 
     & + [B + B'(\alphab)] y + C + C'(\alphab)\} \dot{P}_n(\alphab,
       \lambdab_1; y) \\[0.2cm]  
     & + [(\lambda_{n+1} + \lambda) y + \mu_{n+1} + \mu] P_n(\alphab,
       \lambdab_1; y),  \\[0.2cm]
     \mbox{Class 2:} & P_{n+1}(\alphab, \lambdab; y) = 2y \{A + A'(\alphab)
       + [B + B'(\alphab)] y\} \dot{P}_n(\alphab, \lambdab_1; y) \\[0.2cm] 
     & + \{\lambda_{n+1} + \lambda - n [A + A'(\alphab)] + [\mu_{n+1} + \mu
       - n (B + B'(\alphab))] y\} \\[0.2cm]  
     & \times P_n(\alphab, \lambdab_1; y), \\[0.2cm] 
     \mbox{Class 3:} & P_{n+1}(\alphab, \lambdab; y) = \{[C + C'(\alphab)]
       y + D + D'(\alphab)\} \\[0.2cm]
     & \times \left[ - (A y^2 + B) \dot{P}_n(\alphab, \lambdab_1; y) + nAy
       P_n(\alphab, \lambdab_1; y)\right] \\[0.2cm]
     & + [(\lambda_{n+1} + \lambda) y + \mu_{n+1} + \mu] P_n(\alphab,
        \lambdab_1; y). 
  \end{array}   
\end{equation}

\par
%
%===================================================
%
\section{Some simple examples}

\subsection{Particle in a box and trigonometric P\"oschl-Teller potential}

Let us consider the superpotential
\begin{equation}
  W(\lambda; x) = \lambda \tan x, \qquad - \frac{\pi}{2} \le x \le
  \frac{\pi}{2}.  \label{eq:ex1-W}
\end{equation}
In the undeformed case, for $\lambda = A >1$, it gives rise to the
trigonometric P\"oschl-Teller potential~\cite{sukumar85a}
\begin{equation}
  V_{\rm eff}(A; x) = A(A-1) \sec^2 x,  \label{eq:trig}
\end{equation}
whose bound-state energies and wavefunctions~\cite{cq99} are given by $E_n =
(A+n)^2$ and
$\psi_n(x) \propto (\cos x)^A C^{(A)}_n(\sin x)$, $n=0$, 1,
2,~\ldots. The particle-in-a-box problem being the limiting case
of P\"oschl-Teller for $A \to 1$ corresponds to
\begin{equation}
  V_{\rm eff}(x) = \left\{\begin{array}{ll}
       0 & \mbox{if $ - \frac{\pi}{2} < x < \frac{\pi}{2}$} \\[0.2cm]
       \infty & \mbox{if $ x = \pm \frac{\pi}{2}$}
  \end{array} \right..  \label{eq:box}
\end{equation}
\par
%
%---------------------------------------------------
%
The superpotential (\ref{eq:ex1-W}) belongs to class 1 with $\phi(x) = \tan x$
and $\mu = 0$. On choosing $A=C=1$, $B=0$, $A'(\alphab) = \alpha$, and
$B'(\alphab) = C'(\alphab) = 0$, we get
\begin{equation}
  g(\alpha;x) = \alpha \sin^2 x,  \label{eq:ex1-g}
\end{equation}
which for $-1 < \alpha \ne 0$ leads to a positive-definite deforming function
$f(\alpha;x)$.\par
%
%------------------------------------------------
%
In the particle-in-a-box problem, one easily finds that Eqs.~(\ref{eq:C1})
and (\ref{eq:C2}) are fulfilled provided $\lambda_i = (i+1) (1 + \alpha)$ and 
$\epsilon_i = (2i+1) (1 + \alpha)$, $i=0$, 1, 2,~\ldots. The corresponding
energy eigenvalues and wavefunctions are given by
\begin{eqnarray}
  E_n(\alpha, \lambda) & = & (1 + \alpha) (n+1)^2, \label{eq:ex1-E} \\
  \psi_n(\alpha, \lambda; x) & \propto & \frac{(\cos
      x)^{n+1}}{(1 + \alpha \sin^2 x)^{(n+2)/2}} P_n(\alpha, \lambda; \tan x),
      \label{eq:ex1-wf}
\end{eqnarray}
respectively. Here $P_n(\alpha, \lambda; y)$ satisfies the equation
\begin{eqnarray}
  P_{n+1}(\alpha, \lambda; y) & = & - [1 + (1+\alpha) y^2] \dot{P}_n(\alpha, 
    \lambda_1; y) \nonumber \\
  && \mbox{} + (n+3) (1+\alpha) y P_n(\alpha, \lambda_1; y),  \label{eq:ex1-P}
\end{eqnarray}
where a dot denotes derivative with respect to $y$. For any $n=0$, 1,
2,~\ldots, the wavefunctions (\ref{eq:ex1-wf}) are square integrable and
ensure the Hermiticity of $\pi$. Hence in the presence of deformation
(\ref{eq:ex1-g}), the particle-in-a-box problem still has an infinite number of
bound states making up a quadratic spectrum. As can be checked, 
Eqs.~(\ref{eq:ex1-E}) and (\ref{eq:ex1-wf}) go over to the undeformed energies
and wavefunctions since $P_n(\alpha, \lambda; \tan x)$ becomes proportional to 
$\sec^n x\, C^{(1)}_n(\sin x)$ for $\alpha \to 0$.\par
%
%------------------------------------------------
%
For the trigonometric P\"oschl-Teller potential (\ref{eq:trig}), the results
are similar although more complicated. The energies and associated
wavefunctions are then given by
\begin{eqnarray}
  E_n(\alpha, \lambda) & = & (\lambda+n)^2 - \alpha (\lambda-n^2) \nonumber \\
  & = & \left[\case{1}{2}(\Delta+1) + n\right]^2 + \alpha n(n+1) - \case{1}{4}
        \alpha^2,  \label{eq:ex2-E}\\ 
  \psi_n(\alpha, \lambda; x) & \propto & (\cos x)^{\frac{\lambda}
        {1+\alpha} + n} (1 + \alpha \sin^2 x)^{- \frac{1}{2}\left(\frac{\lambda}{1+\alpha}
         +n+1\right)} \nonumber \\
  && \mbox{} \times P_n(\alpha, \lambda; \tan x),
\label{eq:ex2-wf}
\end{eqnarray}
where $\lambda = \frac{1}{2} (1 + \alpha + \Delta)$,  $\Delta \equiv
\sqrt{(1+\alpha)^2 + 4A(A-1)}$ and 
$P_n(\alpha,
\lambda; y)$ satisfies the equation
\begin{eqnarray}
  P_{n+1}(\alpha, \lambda; y) & = & - [1 + (1+\alpha) y^2] \dot{P}_n(\alpha,
     \lambda_1; y) \nonumber \\
  && \mbox{} + [2\lambda + (n+1) (1+\alpha)] y P_n(\alpha,\lambda_1; y)
     \label{eq:ex2-P}
\end{eqnarray}
with $\lambda_1 = \lambda + 1 + \alpha$. All functions $\psi_n(\alpha, \lambda;
x)$,
$n=0$, 1, 2,~\ldots, are physically acceptable as bound-state
wavefunctions.\par
%
%-----------------------------------------------
%
The starting potential in the PDM Schr\"odinger equation (\ref{eq:PDM-SE}) can
be written as
$V = V_{\rm eff} - \tv$, where $V_{\rm eff}$ is given by (\ref{eq:trig}) or
(\ref{eq:box}), while 
\begin{equation}
  \tv(\alpha, \rho, \sigma; x) = - (\rho + \sigma) \alpha^2 \cos^2 2x + \rho 
  \alpha (2+\alpha) \cos 2x + \sigma \alpha^2. \label{eq:ex1-tV}
\end{equation}
%
%+++++++++++++++++++++++++++++++++++++++++++++++
% 
\subsection{Free particle and hyperbolic P\"oschl-Teller potential}

In the undeformed case, the hyperbolic counterpart of the superpotential
(\ref{eq:ex1-W}), namely
\begin{equation}
  W(\lambda; x) = \lambda \tanh x, \qquad - \infty < x < \infty,
\label{eq:ex3-W}
\end{equation}
corresponds for $\lambda = A > 0$ to the hyperbolic P\"oschl-Teller
potential~\cite{sukumar85b}
\begin{equation}
  V_{\rm eff}(A; x) = - A(A+1) \sech^2 x,
\end{equation}
whose $n_{\rm max} + 1$ bound-state energies and wavefunctions are given by
$E_n = - (A-n)^2$ and $\psi_n(x) \propto (\sech x)^{A-n}
C^{(A-n+\frac{1}{2})}_n(\tanh x)$, $n = 0$ , 1,~\ldots, $n_{\rm max}$, with
$A-1 \le n_{\rm max} < A$. For $A \to 0$, we get the free-particle problem
as a limiting case.\par
%
%--------------------------------------------------------
%
The superpotential (\ref{eq:ex3-W}) also belongs to class 1 with $\phi(x) =
\tanh x$ and $\mu = 0$. On choosing this time $A=-C=-1$, $B=0$, $A'(\alphab) =
\alpha$, and $B'(\alphab) = C'(\alphab) = 0$, we get
\begin{equation}
  g(\alpha;x) = \alpha \sinh^2 x,  \label{eq:ex3-g}
\end{equation}
which for $0 < \alpha < 1$ leads to a positive-definite deforming function
$f(\alpha;x)$.\par
%
%------------------------------------------------
%
In the case of the hyperbolic P\"oschl-Teller potential and $g(\alpha;x)$
given in (\ref{eq:ex3-g}), the energies and associated wavefunctions can be
written as
\begin{eqnarray}
  E_n(\alpha, \lambda) & = & - (\lambda-n)^2 - \alpha (\lambda+n^2) \nonumber
        \\
  & = & - \left[\case{1}{2}(\Delta-1) - n\right]^2 + \alpha n(n+1) +
        \case{1}{4}\alpha^2,  \label{eq:ex3-E}\\ 
  \psi_n(\alpha, \lambda; x) & \propto & (\sech x)^{\frac{\lambda}
        {1-\alpha} - n} (1 + \alpha \sinh^2 x)^{\frac{1}{2}\left(
        \frac{\lambda} {1-\alpha}-n-1\right)} \nonumber \\
  && \mbox{} \times P_n(\alpha, \lambda; \tanh x) \label{eq:ex3-wf}
\end{eqnarray}
where $\lambda = \frac{1}{2}(\alpha - 1 + \Delta)$, $\Delta \equiv
\sqrt{(1-\alpha)^2 + 4A(A+1)}$ and $P_n(\alpha, \lambda; y)$ satisfies an
equation similar to (\ref{eq:ex2-P}). However, it turns out that although for
any $n=0$, 1, 2,~\ldots, $\psi_n(\alpha, \lambda; x)$ is square integrable on
the real line, it does not satisfy the condition
$|\psi|^2 f \to 0$ at the boundaries $x \to \pm \infty$. Hence, with a deformed
function corresponding to (\ref{eq:ex3-g}), the hyperbolic P\"oschl-Teller
potential has no bound state.\par
%
%-------------------------------------------------
%
This result can be extended to the free-particle problem. It contrasts with
what was obtained in~\cite{bagchi04} in another context and illustrates the
strong dependence of the bound-state spectrum on the mass environment.\par
%
%=================================================
%
\section{Results and comments}

We have used the procedure illustrated in Sect.~5 to obtain a deforming
function satisfying Eqs.~(\ref{eq:C1}) and (\ref{eq:C2}), as well as the
resulting bound-state energies and wavefunctions, for all the SI potentials
contained in Table 4.1 of~\cite{cooper}. Below we list the results
obtained for $g$, $E_n$, and $\tv$.\par
%
%----------------------------------------------
%
\medskip
\noindent
{\sl Shifted oscillator:}
\smallskip
\[ V_{\rm eff} = \frac{1}{4} \omega^2 \left(x - \frac{2b}{\omega}\right)^2, \]
\[ g = \alpha x^2 + 2\beta x, \qquad \alpha > \beta^2 \ge 0, \]
\begin{eqnarray*}
  E_n & = & \left(n + \frac{1}{2}\right) \Delta + \left(n^2 + n + \frac{1}{2}\right)
       \alpha + b^2 \\ 
  && \mbox{} - \left(\frac{[(2n+1)\Delta + (2n^2+2n+1)\alpha] \beta -
    b\omega}{\Delta + (2n+1)\alpha}\right)^2, \nonumber \\
  && \Delta \equiv \sqrt{\omega^2 + \alpha^2},  \qquad n=0, 1, 2, \ldots,  
\end{eqnarray*}
\[ \tv = 2(\rho + 2\sigma) \alpha x (\alpha x + 2\beta) + 2\rho\alpha +
    4\sigma \beta^2. \]
\par
%
%----------------------------------------------
%
\medskip
\noindent
{\sl Three-dimensional oscillator:}
\smallskip
\[ V_{\rm eff} = \frac{1}{4} \omega^2 x^2 + \frac{l(l+1)}{x^2}, \qquad 0
   \le x < \infty, \]
\[ g = \alpha x^2, \qquad \alpha > 0, \]
\begin{eqnarray*} 
  E_n & = & \Delta \left(2n + l + \frac{3}{2}\right) + \alpha
      \left[2(n+l+1)(2n+1) + \frac{1}{2}\right], \\
  && \Delta \equiv \sqrt{\omega^2 + \alpha^2}, \qquad n=0, 1, 2, \ldots, 
\end{eqnarray*}
\[ \tv = 2(\rho + 2\sigma) \alpha^2 x^2  + 2\rho\alpha. \]
\par
%
%--------------------------------------------
% 
\medskip
\noindent
{\sl Coulomb:}
\smallskip
\[ V_{\rm eff} = - \frac{e^2}{x} + \frac{l(l+1)}{x^2}, \qquad 0 \le x <
\infty, \]
\[ g = \alpha x,  \qquad \alpha > 0, \]
\begin{eqnarray*} 
  E_n & = & - \left(\frac{e^2 - \alpha[n^2 + (l+1)(2n+1)]}{2(n+l+1)}\right)^2,
       \\ 
  && \mbox{\rm $n=0, 1, \ldots, n_{\rm max}$, where $n_{\rm max}=$ largest
       integer such that} \\
  && \mbox{\rm $n^2 + (l+1)(2n+1) < \frac{e^2}{\alpha}$ if $\alpha <
       \frac{e^2}{l+1}$}, 
\end{eqnarray*}
\[ \tv = \sigma \alpha^2. \]
\par
%
%--------------------------------------------
% 
\medskip
\noindent
{\sl Morse:}
\smallskip
\[ V_{\rm eff} = B^2 e^{-2x} - B (2A+1) e^{-x}, \qquad A, B > 0, \]
\[ g = \alpha e^{-x}, \qquad \alpha > 0, \]
\begin{eqnarray*} 
  E_n & = & - \frac{1}{4} \left(\frac{2B(2A+1) - [(2n+1)\Delta + (2n^2+2n+1)\alpha]}
       {\Delta + (2n+1)\alpha}\right)^2, \\
  && \Delta \equiv \sqrt{4B^2 + \alpha^2}, \mbox{\rm $n=0, 1, \ldots, 
       n_{\rm max}$, where $n_{\rm max} =$}\\ 
  && \mbox{\rm largest integer smaller than $A$ and such that $\alpha <
\alpha_{\rm max}(n_{\rm
       max})$ with} \\
  && \quad \alpha_{\rm max}(0) = \frac{4A(A+1)B}{2A+1}, \\
  && \quad \alpha_{\rm max}(n) = \frac{B(2A+1)(2n^2+2n+1)}{2n^2(n+1)^2} \\
  && \quad - \frac{B(2n+1)
       [(2A+1)^2+4n^2(n+1)^2]^{1/2}}{2n^2(n+1)^2}, \qquad n=1, 2, \ldots, 
\end{eqnarray*}
\[ \tv = (\rho + \sigma) \alpha^2 e^{-2x} + \rho \alpha e^{-x}. \]
\par
%
%-------------------------------------------
%
\medskip
\noindent
{\sl Eckart:}
\smallskip
\begin{eqnarray*} 
  V_{\rm eff} & = & A(A-1) \csch^2 x - 2B \coth x, \qquad A \ge \frac{3}{2},
    \qquad B > A^2, \\
  && 0 \le x < \infty, 
\end{eqnarray*}
\[ g = \alpha e^{-x} \sinh x, \qquad  -2 \le \alpha \ne 0, \]
\begin{eqnarray*}
  E_n & = & - (A+n)^2 - \left(\frac{B - \frac{1}{2}\alpha[(2n+1)A + n^2]}
       {A+n}\right)^2 \\
  && \mbox{}- \alpha[(2n+1)A + n^2], \\
  && \mbox{\rm $n=0, 1, 2, \ldots \ $ if $\alpha = -2$,} \\
  && \mbox{\rm $n=0, 1, \ldots, n_{\rm max} \ $ if $\alpha > -2$, where
       $n_{\rm max} =$ largest integer} \\
  && \mbox{\rm such that $(A+n)^2 < \frac{2B + \alpha A(A-1)}{2 +
       \alpha}$},  
\end{eqnarray*}
\[ \tv = (\rho + \sigma) \alpha^2 e^{-4x} - \rho \alpha (2+\alpha) e^{-2x}. \]
\par
%
%-------------------------------------------
%
\medskip
\noindent
{\sl Scarf I:}
\smallskip
\begin{eqnarray*}
  V_{\rm eff} & = & (B^2 + A^2 - A) \sec^2 x - B (2A-1) \tan x \sec x, \\
  && 0 < B < A-1, \qquad- \frac{\pi}{2} \le x \le \frac{\pi}{2}, 
\end{eqnarray*}
\[ g = \alpha \sin x, \qquad 0 < |\alpha| < 1, \]
\begin{eqnarray*}
  E_n & = & - \frac{1}{4}(2n + 1 + \Delta_+ + \Delta_-)^2 + \alpha \left(n +
    \frac{1}{2} \right)(\Delta_+ - \Delta_-) \\
  && \mbox{} - \alpha^2 \left(n^2 + n + \frac{1}{2}\right), \\
  && \Delta_{\pm} \equiv \sqrt{\frac{1}{4}(1 \mp
    \alpha)^2 + (A \pm B)(A \pm B - 1)}, \qquad n=0, 1, 2, \ldots,
\end{eqnarray*}
\[ \tv = - (\rho + \sigma) \alpha^2 \sin^2 x - \rho\alpha \sin x + \sigma
\alpha^2. \]
\par
%
%-----------------------------------------
%
\medskip
\noindent
{\sl Rosen-Morse I:}
\smallskip
\[ V_{\rm eff} = A(A-1) \csc^2 x + 2B \cot x, \qquad A \ge \frac{3}{2},
    \qquad 0 \le x \le \pi, \]
\[ g = \sin x (\alpha\cos x + \beta\sin x), \qquad \frac{|\alpha|}{2} <
    \sqrt{1+\beta}, \qquad \beta > -1, \]
\begin{eqnarray*} 
  E_n & = & (A+n)^2 - \left(\frac{B + \frac{1}{2}\alpha[(2n+1)A
    + n^2]}{A+n}\right)^2 + \beta[(2n+1)A + n^2], \\ 
  && n=0, 1, 2, \ldots,  
\end{eqnarray*}
\begin{eqnarray*}
  \tv & = & (\rho + \sigma) \left[\frac{1}{2}(\alpha^2 - \beta^2)\cos 4x +
       \alpha\beta \sin 4x\right] \\
  && \mbox{} + \rho (2+\beta) (- \alpha\sin 2x + \beta\cos 2x) + (- \rho +
       \sigma) \frac{1}{2}(\alpha^2 + \beta^2).
\end{eqnarray*}
\par
%
%-------------------------------------------------
%
Three potentials considered in~\cite{cooper} are missing from the list:
Scarf II because  no nontrivial values of the parameters may ensure positive
definiteness of $f(\alphab; x)$, Rosen-Morse II and generalized
P\"oschl-Teller because they do not have any bound state in the deformed case.
\par
%
%-------------------------------------------------
%
{}For the remaining potentials, strikingly distinct influences of deformation 
or mass parameters on bound-state energy spectra are observed. In some cases
(shifted oscillator, three-dimensional oscillator, Scarf I and Rosen-Morse I),
the infinite number of bound states of conventional quantum mechanics remains
infinite after the onset of deformation. Similarly, for Morse potential and
for Eckart potential with $\alpha \ne -2$, one keeps a finite number of bound
states. For the Coulomb potential, however, the infinite number of bound
states is converted into a finite one, while for Eckart potential with $\alpha
= -2$, the finite number of bound states becomes infinite.  It is also
remarkable that, whenever finite, the bound-state number becomes dependent on
the deforming parameter.\par
%
%-----------------------------------------------------
% 
{}For the potentials $V$ to be used in the PDM Schr\"odinger equation
(\ref{eq:PDM-SE}), we get either the same shape as $V_{\rm eff}$ (shifted
oscillator, three-dimensional oscillator, Coulomb and Morse) or a different
shape (remaining potentials). In the first case, the mass and ambiguity
parameters only lead to a renormalization of the potential parameters and/or
an energy shift.\par
%
%==================================================
% 
\section{Conclusion}

In this communication, we have shown how to generate new exactly solvable PDM
(resp.~deformed) Schr\"odinger equations with a bound-state spectrum different
from that of the corresponding constant-mass (resp.~undeformed) Schr\"odinger
equations and we have illustrated our method by several examples. In addition,
we have demonstrated the importance of the Hermiticity condition on the
deformed momentum operator for the existence of bound states.\par
%
%=================================================
%  
\section*{Acknowledgments}

A.\ B.\ thanks the University Grants Commission, New Delhi for the award of a
Junior Research Fellowship. C.\ Q.\ is a Research Director, National Fund for
Scientific Research (FNRS), Belgium.\par

%%%%%%%%%%%%%%%%%%%%%%%%%%%%%%%%%%%%%%%%%%%%%%%%%%%%%%%%%%%%%
% Doing references:                                         %
%%%%%%%%%%%%%%%%%%%%%%%%%%%%%%%%%%%%%%%%%%%%%%%%%%%%%%%%%%%%%

\end{document}